\documentclass[12pt]{iopart}

\usepackage[dvips]{epsfig}
\begin{document}

\title{Conductance oscillation and  quantization in  monoatomic Al wires}

\author{Ying Xu$^{1}$\footnote{yingxu@jxnu.edu.cn}, Xingqiang Shi$^{2}$, Zhi Zeng$^{2}$,
Zhao Yang Zeng$^{1}$, and Baowen Li$^{3,4}$}

\address { $^1$Department of Physics, Jangxi Normal University, Nanchang 330022, China\\
$^2$Key Laboratory of Material Physics, Institute of Solid State
Physics, Chinese Academy of Sciences, Hefei, Anhui 230031, China\\
$^3$ Department of Physics and Centre for Computational Science and
Engineering, National University of Singapore, 117542, Singapore\\
$^4$ NUS Graduate School for Integrative Sciences and Engineering,
Singapore 117597,  Singapore }

\begin{abstract}

We present first-principles calculations for the transport
properties of  monoatomic Al wires sandwiched between Al(100)
electrodes. The conductance of the  monoatomic Al wires oscillates
with the number of the constituent atoms as a function of the wire
length, either with a period of four-atom for wires with the
typical interatomic spacing or a period of six-atom with the
interatomic spacing of the bulk fcc aluminum, indicating a
dependence of the period of conductance oscillation on the
interatomic distance of the monoatomic Al wires.

\end{abstract}

 \vspace{1cm}

\section{Introduction}

Atomic wires, which can be produced by means of either scanning
tunnelling microscope[1], mechanically controllable break junction
techniques[2], or transmission electron microscope[3], have
attracted increasing attentions both theoretically and
experimentally [4]. For monovalent atomic wires, the even-odd
oscillations of the wire conductance with the wire length has been
analyzed explicitly and demonstrated numerically[5],  and perfect
transmission is found for atom chains of an odd number atoms and a
smaller transmission in the even number case.  The even-odd
oscillations of conductance has been confirmed experimentally for
atomic wires formed by noble-metal Au, Pt and Ir atoms[6]. An
anomalous oscillation of conductance is found for atomic chains
composed of alkali-metal atoms (such as Na, Cs) [7-9], with
perfect transmission in the case of even number. However, for
polyvalent atomic wires such as Pb and Al, multiple conductance
channels contribute to the conductance, the situation becomes more
complicated, and more interesting phenomena are expected.

The transport properties of monoatomic wires formed by Al atoms
have been studied with the first-principles method by several
groups[10-12].  Lang
  has investigated the resistance of Al atomic wires connecting
  two semi-infinite metallic electrodes[10].
  Kobayashi et al.[11] has studied the conducting channels for monoatomic Al wires
  by the recursion-transfer matrix
  method.  They reveal that three open channels
can contribute to the current through  the Al atomic wire, and
channel transmission are sensitive to the geometry of the wire.
Ono and Hirose [12] have studied the electronic conductance of a
three-Al-atom suspended between genuine semi-infinite aluminum
crystalline electrodes. They have demonstrated that just one
conducting channel is widely open at the Fermi level. However,  we
notice that, in the above calculations, the interatomic spacing
between Al atoms is assumed to be the same as the bulk fcc one.
The influence of the electrode model structure on the sandwiched
wire has been clarified by Fujimoto et al.[13]. Conductance
oscillations with a period of four-atom have been shown by
Thygesen and Jocobsen with the help of the plane-wave based
pseudopotential code for the  Al wire between two Al(111)
electrodes[14]. The interatomic spacing of the wire is set to be
2.39 \AA, the typical interatomic distance of the Al wires.

In the present paper, we perform a first-principles calculations
on the conductance of the monoatomic Al wires attached to a pair
of Al(100) bulk electrodes so as to clarify the effect of the
interatomic spacing of the wire.  Our results are consistent with
that obtained in ref. [14], when a smaller interatomic atomic
spacing (the typical interatomic distance of Al wires) is used.
While if we choose the interatomic distance of the
 bulk fcc aluminum, the six-atom period of the conductance oscillation is
observed, which is different from the former case. Our results
suggest that the period of conductance oscillation depends on the
interatomic spacing of the wire.

\section{Methodology}
   The calculations have been performed by using a recently developed
first-principles package TranSIESTA-C[15-17]. The package is based
on the combination of density function theory (DFT) implemented in
the well tested SIESTA method with the nonequilibrium Green
function technique. TranSIESTA-C is capable of modelling
self-consistently the electrical properties of nanoscale devices
that consist of an atomic scale system coupled to two
semi-infinite electrodes. The system considered can be divided
into three parts: the left electrode, the scattering region and
the right electrode. The central region also consists of two
layers of surface atoms at left and three layers of surface atoms
at right in order to include the interaction between the electrode
and the atomic wire. The end of Al atomic wire  is fixed at the
hollow sites of the Al(100). The distance between the end atom of
the wire and the electrode edge is set to be 1 \AA. In our
calculations two values of interatomic spacing of the wire are
used, one is 2.86 \AA, the same distance as in the bulk fcc
aluminum; the other is 2.39 \AA, typical Al-Al atom spacing. The
{\sl z}-direction is set to be parallel to the wire axis.  Figure
1 shows the system we considered: a monoatomic Al wire is
sandwiched between a pair of Al(100) electrodes.

\begin{figure}
\begin{center}
\vspace{10mm}
\includegraphics[width=10cm]{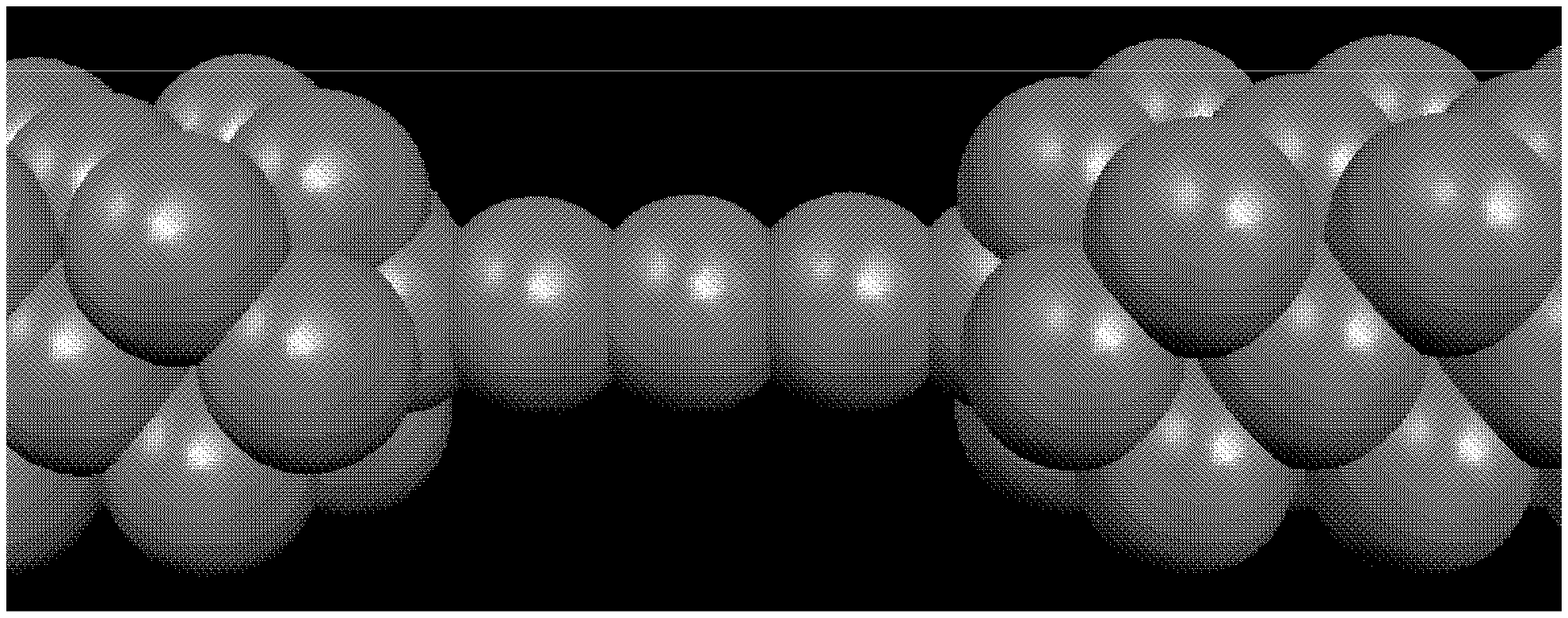}
\caption{A schematic representation of an aluminum atom chain with
5 Al atoms sandwiched between a pair of Al(100) electrodes.}
\end{center}
\end{figure}

\section{Results and discussions}

   We first calculate the conductance of the monoatomic Al wire
   with the typical interatomic spacing 2.39 \AA, as a function
of the number of the constituent aluminum atoms.  The results are
shown Fig. 2 (a). The figure clear shows an oscillation with
 a period of four-atom which is similar to
that one found in Ref. [14]. However, it is worthwhile to point
out that in our calculations, the equilibrium conductance of the
system varies between $0.8\sim1.9$ G$_{0}$ (G$_{0}$=2e$^{2}$/h),
and the maximal conductance occurs as the number of wire atoms
takes values  5, 9, 13 $\cdots$. While in ref. [14], the maximal
values of the conductance correspond to the atom number 3, 7, 11
$\cdots$. and the conductance range is $0.5\sim1.7$ G$_{0}$. The
different phase and amplitude of the conductance oscillations may
originate from the geometrical structure  difference of the
electrodes. We use a pair of semi-infinite Al(100) electrodes
while Thygesen and Jocobsen [14] use two Al(111) electrodes.

One may ask, if the monoatomic Al chain is elongated, namely, the
spacing between Al atoms is increased, can one expect the same
four-atom period  for the conductance as a function of the number
of Al atoms? To answer this question, we calculate the conductance
of Al atom wires coupled to the same pair of Al(100) electrodes as
the number of the Al atoms forming the chains, as the spacing of
the chains is chosen the same as the interatomic spacing of the
bulk fcc aluminum 2.86 \AA. The results are shown in Figure 2 (b).
To our surprise, the conductance exhibits oscillations with a
period of six-atom. Such an observation suggests that the period
of the conductance oscillations of monoatomic Al wires is related
to the interatomic spacing of the wires.

\begin{figure}
\begin{center}
\vspace{30mm}
\includegraphics[width=15cm]{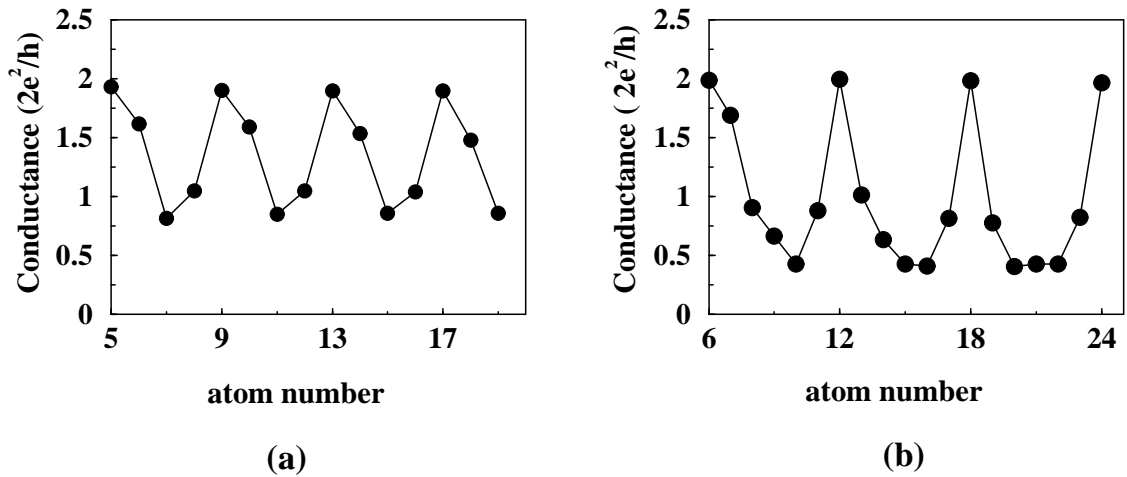}
\caption{Conductance of monoatomic Al wires connecting to a pair
of Al(100) electrode as a function of the number of  Al atoms. (a)
Al interatomic spacing is 2.39 \AA\. (b) The interatomic spacing
is 2.86 \AA\ which is the same as that of the bulk fcc aluminum.}
\end{center}
\end{figure}

To confirm our observation of the conductance oscillation with a
period of four-atom or six-atom for  the monoatomic Al wire with
different interatomic spacings sandwiched between a pair of
Al(100) electrodes, we also calculate the transmission spectra of
the system and show the results in Fig. 3. One can see a clear
level splitting as the increase of the number of the resonant
peaks with increase of the number of the constituent Al atoms.
This can be easily tested from an analysis of the Molecular
Projected Self-consistent Hamiltonian (MPSH). The eigenstates of
MPSH are in fact  the molecular orbitals renormalized by the
molecule-electrode couplings. As an example, one would expect 12
eigen levels for the wire containing 12 monovalent atoms.
According the Landauer-B$\ddot{u}$ttiker theory, the equilibrium
conductance is proportional to the transmission probability at the
Fermi level. Therefore the conductance oscillations of monoatomic
Al wires are equivalent to the coincidence of transmission peaks
at the Fermi level for some specific numbers of the Al atoms
forming the wire. From Figs. 3 (a) and (b), one clearly sees such
a coincidence of transmission peaks as the number of the Al atoms
is 5, 9, 13, 17 for the wire with the  interatomic distance of the
bulk fcc aluminum, and 6, 12,18,24 for the wire with the typical
interatomic distance of Al.

\begin{figure}
\begin{center}
\vspace{10mm}
\includegraphics[width=15cm]{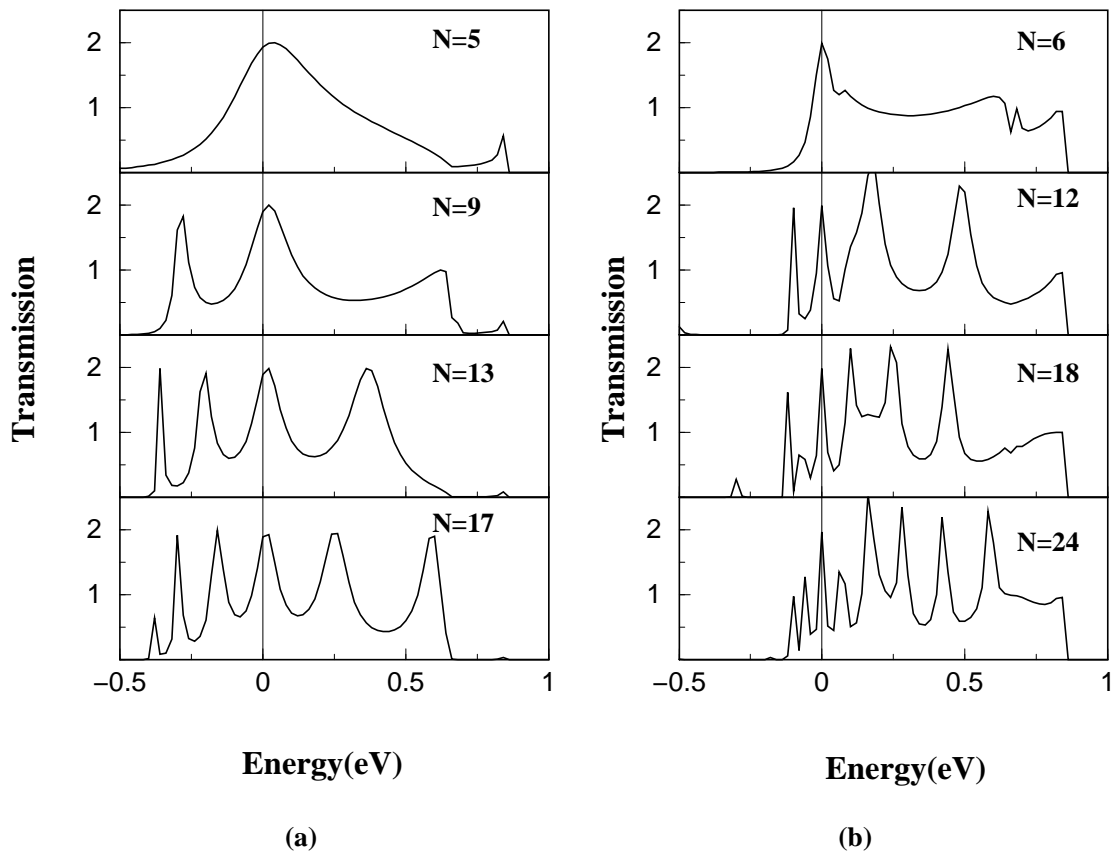}
\caption{Transmission spectra for monoatomic Al wires  for
 different number of constituent Al atoms.  (a) With the typical
 interatomic spacing of Al 2.39 \AA\. (b) With
 the interatomic spacing of the  bulk fcc  aluminum  2.86 \AA\ . }

\end{center}
\end{figure}

  The eigenchannel decomposition of transmission spectra
  provides some useful information about transport properties of
two-probe systems. Therefore, we look at the transmission spectra
of a three-Al-atom wire with the interatomic spacing of the bulk
fcc aluminum 2.86 \AA\. The eigenchannel transmission spectra is
shown in Figure 4. We note that the transmission spectra in this
work are different from the previous results for the three-Al-atom
wire based on various types of jellium  electrode models. It
confirms that the conductance of Al atomic wires is dependent on
the detailed geometric structure of the electrodes[13].

It is well known that, the valence electron configuration of Al
atom is 3{\sl s}$^{2}$3{\sl p}$^{1}$. The {\sl s} and {\sl
p$_{z}$} orbitals constitute $\sigma$-character channel and {\sl
p$_{x}$}, {\sl p$_{y}$} orbitals constitute degenerate
$\pi$-character channels, as clearly shown  in Fig. 4. The
$\sigma$ channel contributes nothing to current and thus the
equilibrium conductance, because Fig. 4 tells us that the
transmission of the $\sigma$ channel at the Fermi level is zero.
Two degenerate $\pi$ channels contribute a small amount to the
equilibrium conductance of the three-Al-atom wire, since the
transmission probabilities of these two degenerate $\pi$ channels
are the same, of the value of about 0.07. We also calculate the
eigenchannel transmission spectra for the monoatomic Al wire with
different numbers  of the  constituent atoms.  We find that for
the Al wires containing 6, 12, 18 Al atoms, two degenerate $\pi$
channels contribute mainly to the equilibrium conductance, each
contribution to conductance being about one conductance quanta one
G$_0$; and the contribution from the $\sigma$ channel is still
zero. Such a conductance quantization is suggested as a result of
charge neutrality and resonant character of the sharp tip
structure of the wire-electrode contacts[18].

\begin{figure}
\begin{center}
\vspace{10mm}
\includegraphics[width=10cm]{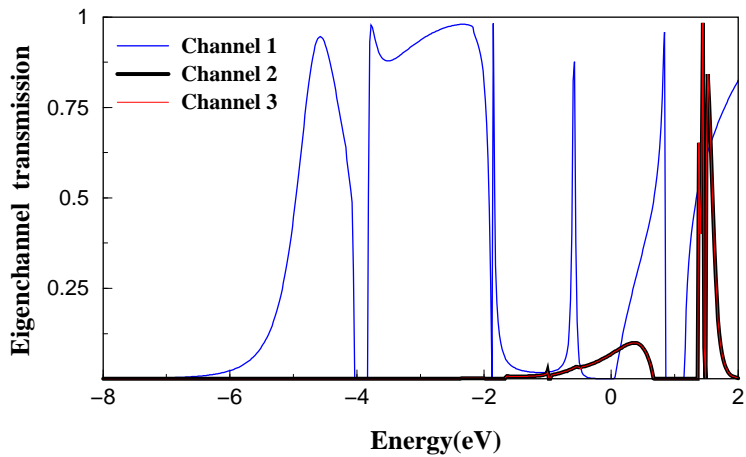}
\caption{Eigenchannel transmission spectra of the $\sigma$ channel
(channel 1) and the two degenerate $\pi$  transmission
 eigenchannels (channel 2 and 3) for the  three-Al-atom wire with the interatomic spacing of the bulk fcc aluminum
 attached  to a pair of Al (100) electrodes.}
\end{center}
\end{figure}

  \begin{figure}
\begin{center}
\vspace{30mm}
\includegraphics[width=15cm]{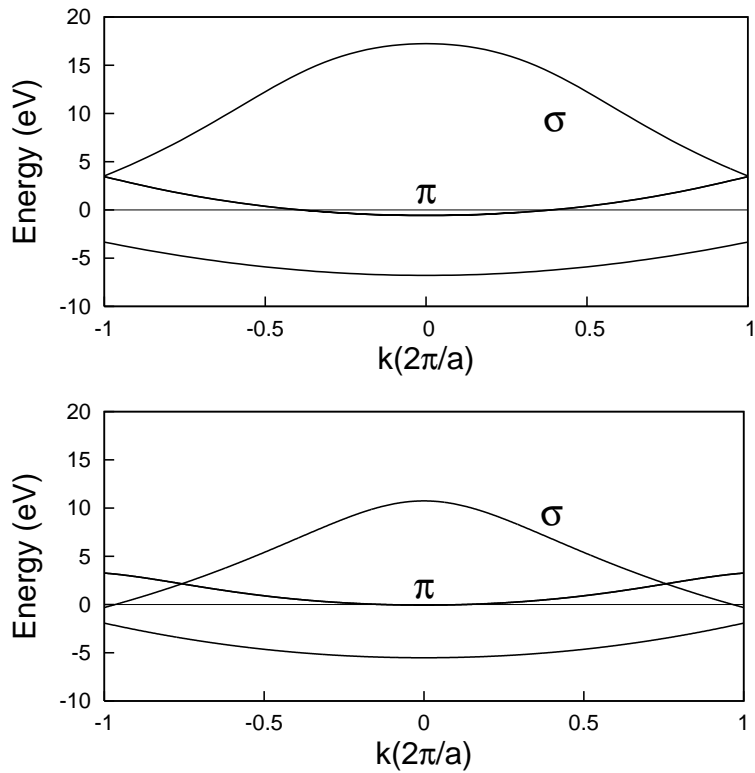}
\caption{The energy band structure of infinite Al wires with the
typical interatomic distance 2.39 \AA\ (upper panel) and the
interatomic distance of the bulk fcc aluminum 2.86 \AA\ (lower
panel) }
\end{center}
\end{figure}

To explain the six-atom period of the conductance oscillation of
monoatomic Al wires with the interatomic spacing of the bulk fcc
aluminum, we perform band structure calculations for an infinite
Al wire with the typical interatomic distance 2.39 \AA\ and the
interatomic distance of the bulk fcc  aluminum  2.86 \AA\ attached
to a pair of Al(100) electrodes. The results are shown in Figure
5. The period of conductance oscillations is determined by the
filling factor of
 the conducting bands with the consideration of local charge
neutrality[14]: the filling factor is the inverse of the period of
the conductance oscillation. Comparing the band structures of the
infinite Al wire with the typical interatomic spacing and with the
 the interatomic spacing of the bulk fcc aluminum, we find that the valence band
  - the bonding $\sigma$ band, formed by the $\sigma$ orbitals of Al
  atoms, is fully filled in both cases. For the infinite Al wires with the typical
 interatomic spacing, the conducting band is formed by the degenerate $\pi$ orbitals,
 while the conducting band includes the contributions of both the
 anti-bonding $\sigma$ band and the degenerate $\pi$ band for the
 monoatomic Al wires with the interatomic distance of bulk fcc
aluminum. Such a dependence of the energy band structure of an
infinite Al wire on the interatomic distance may be associated
with the Pierce distortion effect, and is similar to the findings
by Okano et al.[19]. It is known that each Al atom provides three
electrons with two occupying the bonding $\sigma$ valence band and
another occupying the conducting band. Therefore one finds, for an
infinite monoatomic Al wires,
 a filling factor $1/4$  with the typical interatomic distance and
a filling factor $1/6$ with the interatomic distance of the bulk
fcc aluminum. It is such a filling factor imposed by local charge
neutrality that causes the conductance oscillation with a period
of either four-atom or six-atom, depending on the interatomic
distance of Al atoms forming the wire. However, it is difficult to
determine the phase of the conductance oscillations of the
monoatomic Al wires, due to the limitations of the soft package
TranSIESTA-C we have used.

\section{Conclusion}
    In conclusion, we have investigated conductance oscillations of a monoatomic
Al wires with different interatomic spacings sandwiched between a
pair of Al (100) electrodes with a well-developed soft package
TranSIESTA-C. Conductance oscillations with a period of four-atom
and six-atom are observed for the monoatomic wire with the typical
interatomic distance and the interatomic distance of the bulk fcc
aluminum. The period of the conductance oscillations is determined
by the filling factor of the conducting band of Al wires with
specific interatomic spacing.

\section{Acknowledgements}
  This work was supported by the National Natural Science Foundation of China under
 Grant No. 10404010,  the Project-sponsored by SRF for ROCS, SEM, the Foundation of Jiangxi Educational
Committee under  Grant No. 112[2006], and the Talent Fund of
Jiangxi Normal University under Grant No. 1186 and 1187. BL was
supported in part by a FRG grant of NUS.

\end{document}